\documentclass[runningheads,a4paper]{llncs}

\usepackage{graphicx}
\usepackage{booktabs}
\usepackage{amssymb}
\usepackage{amsmath}
\usepackage{algorithm}
\usepackage{algorithmic}
\usepackage{booktabs}
\usepackage{multirow,bigstrut}
\usepackage{color}
\usepackage{subfigure}
\usepackage{url}
\urldef{\mailsa}\path|{kongsb09@mails., lingfeng@}tsinghua.edu.cn|
\urldef{\mailsb}\path|{qmei, zhezhao}@umich.edu|
\urldef{\mailsc}\path|feiye08@gmail.com|
\newcommand{\keywords}[1]{\par\addvspace\baselineskip
\noindent\keywordname\enspace\ignorespaces#1}

\begin{document}


\title{On the Real-time Prediction \\Problems of Bursting Hashtags in Twitter}

%
%
%
\author{Shoubin Kong$^{1}$, Qiaozhu Mei$^{2}$, Ling Feng$^{1}$, Zhe Zhao$^{2}$, Fei Ye$^{3}$
}
%

\institute{
$^{1}$\mailsa\\
$^{2}$\mailsb\\
$^{3}$\mailsc\\
}

%
%

\maketitle

\begin{abstract}
Hundreds of thousands of hashtags are generated every day on Twitter. Only a few become bursting topics. Among the few, only some can be predicted in real-time. In this paper, we take the initiative to conduct a systematic study of a series of challenging real-time prediction problems of bursting hashtags. Which hashtags will become bursting? If they do, when will the burst happen? How long will they remain active? And how soon will they fade away? Based on empirical analysis of real data from Twitter, we provide insightful statistics to answer these questions, which span over the entire lifecycles of hashtags.

\keywords{hashtag, burstiness, real-time prediction}
\end{abstract}

\section{Introduction}

As one of the leading platforms of social communications and information dissemination, Twitter has become a major source of information for common Web users.
An overload of information is being diffused in real-time, which makes it easy for the users to obtain broad perspectives and quick updates about real world events, and in the meantime, makes it difficult for the users to filter useful and trending information from the noisy context. 

Conversations on Twitter are featured with their ``burstiness'', the phenomenon that a topic of discussion suddenly gains a considerable popularity, and then quickly fades away. Such bursting topics are usually triggered by breaking news, real world events, malicious rumors, or various types of behavior cascades such as campaigns of persuasion.

These bursting topics, usually referred to as trending topics, provide users with fresh discoveries and timely updates of events. Much study has also investigated the value of the bursting topics in a broader context. Bursts of topics, sentiments, and questions have been demonstrated to have a predictive power of product sales \cite{gruhl2005predictive}, stock market \cite{bollen2011twitter}, search engine queries \cite{zhao2013questions}, outburst of diseases \cite{ritterman2009using}, elections \cite{tumasjan2011election}, and even natural disasters \cite{sakaki2010earthquake}. Therefore, an earlier detection of such trending topics implies an increased revenue, a reduced damage, a timely treatment, and better decision-making in general. To help people discover the bursting topics in time, twitter deploys a list of trending topics as long as they are detected.

However, it may be already too late to react even if a burst can be \textit{detected} in no time. On April 23rd, 2013, a false claim about explosions at the White House and the injury of the president sent by the hacked account of the Associated Press quickly became an explosive burst on Twitter \footnote{\url{http://www.foxnews.com/us/2013/04/23/hackers-break-into-associated-press-twitter-account/}}. Although the rumor was debunked and the hacked account was deleted as soon as the burst was detected, damage had been made - the bursting topic had shaken the stock market so badly that the Dow Jones Indices experienced a sudden drop of more than 100. If only we can \textit{predict} the outbreak of a topic before it bursts! But can we?

Hashtags, user-specified strings starting with a \# symbol, have been commonly used as identities of topics in Twitter. From a 10\% random sample of the Tweet stream, we can identify about 400,000 new hashtags every day. However, only dozens of them become bursting. Among the dozens, there may be an even smaller proportion which one can predict in real-time. What are the proportions? How effective is the prediction, and what are the most important factors? How early can the prediction be done?
In this study, we conduct the first \textit{systematic} study of the real-time prediction of bursting hashtags. The key contributions of this paper include the following: 


\begin{enumerate}
\item We take the initiative to provide formal definitions of a bursting hashtag as well as three key states in the lifecycle of a bursting hashtag. We define a series of real-time prediction problems that are concerned with these states of bursting hashtags. 

\item We conduct a systematic study of these real-time prediction tasks, by exploring different solutions and different types of features, in particular novel time series features. We provide a comprehensive summary of the distribution of bursting hashtags and the effectiveness of real-time prediction.

\item Experiments are conducted on real datasets from Twitter to evaluate the performance of the proposed solutions. We also experimentally examine effectiveness of different features and analyze their contributions to the prediction performance.
\end{enumerate}


\begin{figure}[!b]
  \centering
  \includegraphics[width=7.4cm, height=5.2cm]{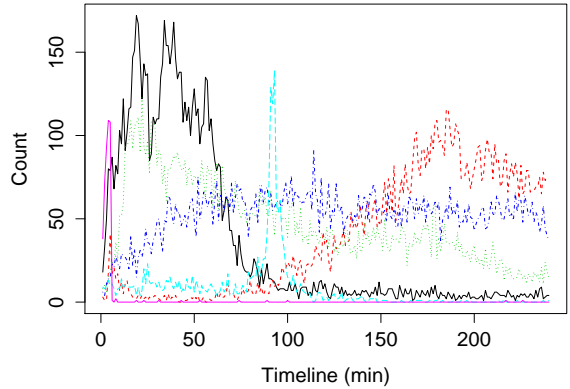}
  \caption{Examples of bursting hashtags}
  \label{fig:bhtag}
\end{figure}

\section{Problem Setup}

\subsection{Definitions}
The lifecycle of a hashtag can be formed as a time series $<c_1,c_2,...,c_t,...>$. $c_t$ denotes the count of tweets containing the hashtag at the $t$-th time interval. Considering the real-time characteristic of Twitter, the granularity of the time interval is set to 1 minute in this study.
Definitions of bursting hashtags are as follows:

\begin{definition}
Prediction-Trigger. A clear majority of hashtags will never get burst, and a substantial number of bursting hashtags have a long dormant period before they burst. The average time before a hashtag gets burst is about \textbf{8.72 days} since the hashtag first appears. Therefore 
we define a trigger to obtain a candidate set of hashtags to be predicted. For each hashtag, a five-minute sliding window is used to check the total count of tweets containing the hashtag within the consecutive five minutes, denoted as $C_{slw}$. If $C_{slw} > \delta$, the prediction is triggered.
\end{definition}

\begin{definition}
Burst. We define the burst of a hashtag by referencing to the definition of spikes in~\cite{gruhl2005predictive}. Within 24 hours since the prediction was triggered, if $c_t$ is greater than max($c_1+\delta,1.5c_1$), $t$ is defined as the onset of burst. 
$\delta$, can be adjusted according to the statistics of real data. We have mentioned that in our dataset about 400,000 new hashtags can be identified every day. $\delta$ is set to 50 in this paper, which makes the ratio of bursting hashtags about 0.6\%\%, i.e., about 25 bursting hashtags can be found each day. If a larger ratio is required, the value of $\delta$ should be set smaller, and vice versa.
\end{definition}

\begin{definition}
Off-Burst. Starting from $c_{t'}$, if all the values are smaller than max($c_1+\delta,1.5c_1$) in the following 24 hours, $t'$ is defined as the end of the burst. We can say the hashtag is off-burst since $t'$.
\end{definition}

\begin{definition}
Death. The definition of ``Off-Burst'' corresponds to the definition of ``Burst''. Analogously, the ``Death'' is defined corresponding to the definition of ``Prediction-Trigger''. If a bursting hashtag could no longer satisfy the condition for triggering prediction in consecutive 24 hours, the hashtag is considered dead. In other words, a complete lifecycle of the bursting hashtag come to an end.
\end{definition}

\begin{figure}[!t]
  \centering
  \includegraphics[width=4.5cm, height=8cm]{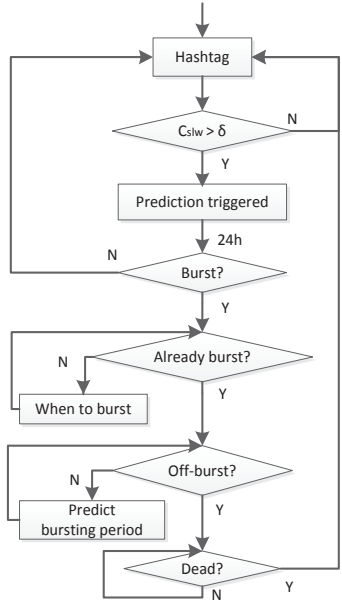}
  \caption{The framework of real-time prediction}
  \label{fig:fpp}
  \vspace*{-12pt}
\end{figure}

Fig.~\ref{fig:bhtag} shows several examples of bursting hashtags. It can be observed that they vary in when they burst and how long the bursting is sustained. Based on the definitions above, we propose a framework of the real-time prediction shown in Fig.~\ref{fig:fpp}, covering the entire lifecycles of hashtags. When a new hashtag comes, a five-minute sliding window is used to constantly check whether it triggers prediction. When it satisfies the triggering condition, real-time prediction is triggered. The first prediction task is to predict whether it will be a bursting hashtag. If it will be, then we predict when it will burst, i.e., the \textbf{T}ime period \textbf{B}efore the onset of \textbf{B}urst ($TBB$). For a hashtag that has already burst, we skip the first two prediction tasks and directly predict when it will be off-burst, i.e., the \textbf{T}ime period that it can \textbf{R}emain \textbf{A}ctive ($TRA$). When this bursting hashtag is dead, it is taken as a new hashtag, entering the prediction process again. In other words, when a hashtag comes to the end of last lifecycle, it automatically starts next round of life.
\subsection{Prediction Tasks and Solutions}
Four prediction problems have been raised over the lifecycle of a hashtag. In this study, we focus on the first three problems closely related to ``burst".

Task 1. Will a hashtag be a bursting hashtag? This problem can be framed to a normal binary classification task.

Input: A set of candidate hashtags which triggered prediction $HT = \{ht_1, ht_2, ...\}$ .

Output: A class label for each hashtag $L(h_i), h_i\in HT$, indicating whether it will be a bursting one.
\begin{algorithm}[!ht]
\caption{ Training optimal classification model. }
\label{alg:wsvm}
\begin{algorithmic}[1]
\REQUIRE ~~\\
    $TNS$: training set; $TTS$: training-test set; \\
    $PNR$: the ratio of negative samples to positive samples; \\
    $R_w=\{1,2,...,2PNR\}$: the range of the weight for positive class;
\ENSURE ~~\\
    $C_{opt}$: the optimal classifier;
\STATE $F_{max} \leftarrow 0$;
\FORALL {$w\in R_w$}
    \STATE Training a weighted SVM classifier $C_w$ on $TNS$;
    \STATE Compute $F_1$-score $F^w_1$ by applying $C_w$ to $TTS$;
    \IF  {$F^w_1 > F_{max}$}
        \STATE $C_{opt} \leftarrow C_w$;
        \STATE $F_{max} \leftarrow F^w_1$;
    \ENDIF
\ENDFOR
\RETURN $C_{opt}$;
\end{algorithmic}
\end{algorithm}

Solution: We propose a weighted SVM-based method to solve this problem, whose dataset is unbalanced. An optimal weight for the positive class is needed to train the classification model. Algorithm~\ref{alg:wsvm} shows the process of optimizing the weight for the positive class. Since the dataset is unbalanced, $F_1$-score is used as the criteria for training the optimal model. At the same time, we also tried several related methods to evaluate the performance. The evaluation results are demonstrated in Section 4.2.

Task 2. If a hashtag will be a bursting one, when will it get burst? This problem can be framed to a regression task.

Input: A set of bursting hashtags which haven't burst $HB = \{hb_1, hb_2, ...\}$.

Output: The time period (minutes) before the onset of each bursting hashtag $TBB(h_i), h_i\in HB$. Note that predicting the exact value of $TBB$ is extremely difficult and is often not necessary. Therefore we relax the problem and predict the natural logarithm of $TBB$, $log(TBB)$. In other words, we turn to predict the range of the time period.

Solution: We tried five different models to solve this problem, including Linear Regression(LR), Classification And Regression Tree(CART), Gaussian Process Regression(GPR), Support Vector Regression(SVR) and Neural Network(NN). The evaluation results can be found in Section 4.3. 

Task 3. Once a hashtag get burst, how long will it remain active, i.e, when will it be off-burst? This problem can also be framed to a regression task.

Input: A set of bursting hashtags which have got burst $HB^{\prime} = \{hb^{\prime}_1, hb^{\prime}_2, ...\}$.

Output: The time period (minutes) that each bursting hashtag can remain active $TRA(h_i), h_i\in HB^{\prime}$. Similar to task 2, we turn to predict the natural logarithm of $TRA$, $log(TRA)$.

Solution: To solve this problem, we also tried the five different models used in Task 2. The evaluation results are shown in Section 4.4.

\subsection{Statistics and Challenges}
$\delta$, the parameter in the definition of \textquotedblleft burst", can be adjusted according to the statistics of real data. We analyzed a two-month dataset from Nov 1, 2012, to Dec 31, 2012 and found that about 400,000 new hashtags are generated every day. 
Table.~\ref{tab:stc} shows the distribution of bursting hashtags over time. The three keys in the table, $RAB$, $ROB$, and $RAD$, are ratios defined as follows:
\begin{displaymath}
    RAB = \frac{\#hashtags\_already\_burst}{\#bursting\_hashtags}
\end{displaymath}
\begin{displaymath}
    ROB = \frac{\#hashtags\_offburst}{\#bursting\_hashtags}
\end{displaymath}
\begin{displaymath}
    RAD = \frac{\#hashtags\_already\_dead}{\#bursting\_hashtags}
\end{displaymath}
From Table.~\ref{tab:stc} we can obtain three observations. Since the time when the prediction was triggered, about 95\% of bursting hashtags get burst within 6 hours; about 96\% of bursting hashtags are off-burst within 24 hours; about 98\% of bursting hashtags are dead within 48 hours.

The most challenging problem for bursting hashtag prediction comes from the unbalanced data. Table~\ref{tab:trig} shows the distribution of hashtags triggering prediction. It can be seen that, as time goes by the data becomes more and more skewed. The proportion of bursting hashtags in the dataset even goes down to \textbf{0.8\%} at the 6th hour. It is quite challenging to precisely predict so few bursting hashtags from the data set.
\begin{table}[!t]
  \centering
  \caption{Distribution of bursting hashtags over time}
    \begin{tabular}{|c|c|c|c|c|c|c|c|c|}
    \hline
          & 5min  & 15min & 30min & 1h    & 3h    & 6h    & 24h   & 48h \bigstrut\\
    \hline
    RAB   & 30.32\% & 51.02\% & 68.03\% & 80.88\% & 92.68\% & 95.12\% & 100.00\% & 100.00\% \bigstrut\\
    \hline
    ROB   & 4.35\% & 39.53\% & 47.29\% & 59.90\% & 81.15\% & 88.50\% & 96.11\% & 99.60\% \bigstrut\\
    \hline
    RAD   & 0.43\% & 14.60\% & 19.38\% & 32.84\% & 61.95\% & 73.46\% & 88.60\% & 98.15\% \bigstrut\\
    \hline
    \end{tabular}%
  \label{tab:stc}%
\end{table}%
\begin{table}[!t]
  \centering
  \caption{Proportion of bursting hashtags in the dataset}
    \begin{tabular}{|c|c|c|c|c|c|}
    \hline
    5min  & 15min & 30min & 1h    & 3h    & 6h \bigstrut\\
    \hline
    12.39\% & 9.24\% & 6.27\% & 3.58\% & 1.27\% & 0.80\% \bigstrut\\
    \hline
    \end{tabular}%
  \label{tab:trig}%
\end{table}%

\section{Feature Space}
In this section, we explore different types of features which may indicate the future trend of hashtags, including meme features, user features, content features, network features, hashtag features, time series features, and prototype features.
\subsection{Meme Features}
Tweet count. We use the number of tweets containing a hashtag to represent current popularity of the hashtag, instead of using the appearance count of the hashtag. This is because some tweets may use the same hashtag multiple times.

Author count. Besides tweet count for a hashtag, we also consider the unique number of authors who posted tweets containing the hashtag. This feature can be used to recognize those hashtags automatically posted by some fake accounts.

Retweet count. Retweeting is the typical way of information diffusion in Twitter. Interesting information can spread quickly and broadly through retweets. If a user retweeted a tweet, that means the content of the tweet successfully attracted the attention of this user and motivated him to share it. Besides indicating the interestingness of messages, the retweeting behavior of a user may also affect his followers.

Mention count. Mention is a directional sharing behavior in Twitter. Messages can be shared to a designated user using @ as the prefix of the user's name. If a user was mentioned in a tweet with a hashtag, he probably took part in the topic, especially when this mention came from his friends.

Url ratio. A url in Twitter can be a link of a picture, a song, a video, or a piece of news. High ratio of tweets with urls may indicate a topic about a good song, an interesting picture or video, or a piece of breaking news. For example, \#GoodLife, a hashtag with a high url ratio, was about a great new song posted by the hippop musician Lyinheart on Memorial Day.

We also consider the ratio version of author count, retweet count and mention count.

\subsection{User Features}
Total follower count. In Twitter, if a user posts a tweet, this tweet will be shown on the personal pages of the user's followers. When any follower see this tweet(suppose it has a hashtag), he may retweet it or post a new tweet using this hashtag if he is interested in this topic. Therefore, the total count of followers seems to be the potential scale of future adoption of the hashtag.

Maximum follower count. Within the users who adopted a hashtag, if there is one whose follower count is much larger than others, e.g. a celebrity, the followers of this user may play a leading role in the potential adoption of this hashtag. Besides, for two hashtags which have the same total follower count, the maximum follower count may break the tie.

Passivity. Active users often post or retweet tweets following some hashtags. On the contrary, passive users rarely do so unless the topics are attractive enough. The passivity of a user is defined as the reciprocal of average number of tweets posted by this user per day, which is formed as:
\begin{displaymath}
    Psv(u_i) = \frac{N_d(u_i)}{1.0+N_t(u_i)}
\end{displaymath}
where denotes the number of days since the user account was created, and denotes the total number of tweets posted by this user.

\subsection{Content Features}
Special signals. Casual language is commonly used in Twitter. Users often use repeated letters in words to strengthen the mood, for example, goooooood, pleasssssse. Besides, repeated punctuation marks can also indicate users' emotion strength. Repeated exclamation mark (!!!) and question mark (???) are also considered as special signals in this study. The tweets with special signals are counted and used as a feature.

Word-level sentiment strength. For different kinds of hashtags, users tends to use words of varying sentiment strength in tweets. Whether for positive or negative sentiment, those words of strong sentiment usually mean bursting hashtags. For example, in those tweets containing the bursting hashtag \#songsiwillalwayslove, some words of strong positive sentiment were used, such as beautiful, awesome, amazing, favorite, etc. While in tweets containing the bursting hashtag \#LondonRiots, some words of strong negative sentiment were used, such as terrible, sad, sorry, etc. Using SentiWordNet~\cite{baccianella2010sentiwordnet}, a lexical resource for supporting sentiment classification and opinion mining applications, average positive score of words can be computed for each hashtag, as well as average negative score. These two scores are used as features indicating word-level sentiment strength.

Emoticon count. An emoticon is a metacommunicative pictorial representation of a facial expression, usually constructed by punctuation marks or traditional alphabetic. Emoticons are commonly used to express a person's feelings in social media, which fall into two typical types. One type is happy/winking emoticons, the other type is sad/disappointed emoticons. We adopted the collections of emoticons in~\cite{owoputi2013improved}, which focus on Twitter part-of-speech tagging. Here are some examples of emoticons.
\begin{itemize}
\item Happy emoticon:\quad :)\quad  :-)\quad  :')\quad  :]\quad  =] 
\item Sad   emoticon:\quad :(\quad  :-(\quad  :'(\quad  :[\quad  =[ 
\end{itemize}
We count the tweets with happy emoticons and the tweets with sad emoticons separately, and use them as two independent features.

\subsection{Network Features}
As mentioned above, retweets and mentions can accelerate the diffusion of a hashtag. A user network for a hashtag can be constructed by use of retweets and mentions, which is a directed graph G = (V,E). Then we can extract several features from this retweet-mention network.

The order of the graph. The order of the graph is $|V|$, i.e. the number of vertices. Each vertex represents a user involved in the retweet-mention network.

Density. Density of the graph is defined as the total number of edges divided by the total number of possible edges, which is used to describe the general level of connectedness in the graph:
\begin{displaymath}
    Density = \frac{|E|}{|V|\times(|V|-1)}
\end{displaymath}

Average degree. Average degree of the graph is another feature measuring the connectedness in the graph, which is formed as:
\begin{displaymath}
    AveDegree = \frac{2|E|}{|V|}
\end{displaymath}

Entropy of degree distribution. Entropy of degree distribution can be used to measure the heterogeneity of the network, which is defined as:
\begin{displaymath}
    Entropy = - \sum\limits_{k=1}^{|V|}{p(k)log(p(k))}
\end{displaymath}
where $p(k)$ is computed using frequencies of node degree.
\subsection{Hashtag Features}
Length of the hashtag. The length of a hashtag is defined as the number of characters in the hashtag. For example, the length of \#3peopleulove is 12. A hashtag which is too short or too long tends not to become a bursting hashtag.

Case-sensitive hashtag count. The case of a hashtag may be changed during diffusion. For example, \#3peopleulove, \#3PeopleuLove and \#3PeopleULove are the same hashtag. Popular hashtags tend to have more versions of different cases.

Co-occurrence times with other hashtags. Sometimes, some hashtags are not used individually, but are used together with other hashtags, e.g. \#boston\#explosion. Here the co-occurrence times is calculated by the number of tweets with two or more hashtags, one of which is the hashtag to be predicted.
\begin{table*}[!ht]
\newcommand{\tabincell}[2]{\begin{tabular}{@{}#1@{}}#2\end{tabular}}
\newcommand{\pp}{\raisebox{1.5pt}{\scalebox{0.7}{++}} }
  \centering
  \caption{Derivative features from time series}
    \begin{tabular}{|c|c|c|}
    \hline
    Feature Name & Mathematical Presentation & Description \bigstrut\\
    \hline
    mean\_value & $E(c)=1/(t_p-s+1) \times \sum\limits_{j=s}^{t_p}{c_j}$ & \tabincell{c}{mean value of \\the time series} \bigstrut\\
    \hline
    std\_value  & $\sqrt{1/(t_p-s+1) \times \sum\limits_{j=s}^{t_p}{(c_j-E(c))^2}}$ & \tabincell{c}{standard deviation \\of the time series} \bigstrut\\
    \hline
    d\_last\_first & $c_{t_p}-c_s$ & \tabincell{c}{d-value between the last \\point and the first one} \bigstrut\\
    \hline
    d\_last\_max & $c_{t_p}-max(c_j),j\in \{s,s+1,...,t_p\}$ & \tabincell{c}{d-value between the last \\point and the maximum one} \bigstrut\\
    \hline
    d\_last\_min & $c_{t_p}-min(c_j),j\in \{s,s+1,...,t_p\}$ & \tabincell{c}{d-value between the last \\point and the minimum one} \bigstrut\\
    \hline
    idx\_max & $m|c_m=max(c_j),j\in \{s,s+1,...,t_p\}$ & index of the maximum point \bigstrut\\
    \hline
    mean\_fod & $E(fod)=1/(t_p-s) \times \sum\limits_{j=s}^{t_p-1}{|c_{j+1}-c_j|}$ & \tabincell{c}{mean value of the \\absolute first-order derivative} \bigstrut\\
    \hline
    std\_fod  & $\sqrt{1/(t_p-s) \times \sum\limits_{j=s}^{t_p-1}{(|c_{j+1}-c_j|-E(fod))^2}}$ & \tabincell{c}{standard deviation of the \\absolute first-order derivative} \bigstrut\\
    \hline
    last\_fod & $c_{t_p}-c_{t_p-1}$ & \tabincell{c}{last value of the \\first-order derivative} \bigstrut\\
    \hline
    max\_fod & $max(c_{j+1}-c_j),j\in \{s,s+1,...,t_p-1\}$ & \tabincell{c}{maximum value of the \\first-order derivative} \bigstrut\\
    \hline
    d\_pfod\_nfod & \tabincell{c}{$pfod - nfod$; \\if $c_{j+1} \geq c_j$, $pfod$\pp; else $nfod$\pp} & \tabincell{c}{d-value between positive and \\negative first-order derivative} \bigstrut\\
    \hline
    \end{tabular}%
  \label{tab:dfcts}%
\end{table*}%
\subsection{Time Series Features}
Dormant period. Dormant period refers to the time period before the prediction was triggered. For different hashtags, the dormant periods vary from several seconds to several weeks. When predicting how long a bursting hashtag will remain active, a similar feature of how long this hashtag has been bursting is also considered.

Polynomial coefficients. Time series can reflect the revolution of a hashtag. Since the granularity of the time series is 1 minute, sometimes the length of the complete time series is too long, which is unreasonable and unnecessary to be directly used as a feature. We adopted two methods to represent the shape of time series. One is polynomial curve fitting, whose function is formed as:
\begin{displaymath}
    f(x,W) = \sum\limits_{k=0}^{\beta}{w_kx^k},
\end{displaymath}
subject to
\begin{eqnarray}
\notag
\beta =
\begin{cases}
t_p-1   &\mbox{if $t_p\leq6$} \\
6    &\mbox{others.} 
\end{cases}
\end{eqnarray}
where $\beta$ denotes the order of the polynomial function, and $t_p$ denotes the time (by minute) since the prediction was triggered. The upper bound of $\beta$ is set to 6, because larger values lead to over-fitting. $W$, coefficients of polynomial curve fitting is used as the features representing the shape of time series.

Symbolic sequences. The other method to recognize the shape of time series is symbolic representation by use of SAX~\cite{lin2003symbolic}, a state-of-the-art method in time series analysis. SAX is leveraged to reduce the complete time series to a symbolic sequence, e.g., ACBF. Then 3-grams can be generated from the sequence. Note that 3-grams here are different from traditional ones at two aspects. First, the last item must be included, as it represents the latest status at that moment. Second, 3-grams here needn't be continuous. Therefore the 3-grams for the symbolic sequence ACBF are \{ACF, ABF, CBF\}. We can get a ranking list of 3-grams by processing the time series from bursting hashtags in the training set. The top-5 3-grams are used as features.

In addition, some derivative features are defined to describe the characteristics of the time series. Assume the time series obtained is formed as $<c_{s},c_{s+1},c_{s+2},...,c_{t_p}>$, where $s$ denotes the starting minute when the prediction was triggered, and $t_p$ denotes that minute when making a real-time prediction. Then details of the derivative features from time series are given in Table~\ref{tab:dfcts}.

\subsection{Prototype Features}
Prototypes here refer to the similar historic hashtags to the new one to be predicted. When searching prototypes from historic hashtags, all the features but symbolic sequences are used as the representation of hashtags. Polynomial coefficients and symbolic sequences are both features representing the shape of time series. It would be redundant to use both of them. Symbolic sequences are excluded because the experimental results show that it's more effective to use polynomial coefficients. Based on Euclidian Distance, the similarity between hashtags is defined as follows:
\begin{equation}
\begin{aligned}
Sim(h_i, h_j)=\frac{1}{1+\sqrt{\sum\limits_{n=1}^{\alpha}{(F^i_n-F^j_n)^2}}},
\end{aligned}
\end{equation}
where $h_i = <F^i_1, F^i_2, ..., F^i_{\alpha}>$, $h_j = <F^j_1, F^j_2, ..., F^j_{\alpha}>$.

Prototype features are different for the three prediction tasks defined in section 2. For prediction task 1, top-\emph{k} prototypes for the hashtag to be predicted can be found from the historic dataset. In these top-$k$ prototypes, the number of bursting hashtags is used as a feature.
While for prediction task 2 and 3, top-\emph{k} historic \textbf{bursting} prototypes should be found, whose weighted average $TBB$ or $TRA$ is used as a feature. Taking prediction task 2 as an example, the prototype feature ($PF$) is formed as:
\begin{equation}
\label{for:pf}
{PF(h_p) = \frac{\sum\limits_{i=1}^{k}{Sim(h_i,h_p) \times TBB(h_i)}}{\sum\limits_{i=1}^{k}{Sim(h_i,h_p)}}}.
\end{equation}
Considering $k$ from 1 to 10, we can get 10 prototype features.


\begin{table}[!b]
\newcommand{\tabincell}[2]{\begin{tabular}{@{}#1@{}}#2\end{tabular}}
  \centering
  \caption{Datasets}
    \begin{tabular}{|c|c|c|c|}
    \hline
    Dataset & Time Period & \tabincell{c}{\#Positive \\Samples} & \tabincell{c}{\#Negative \\Samples} \bigstrut\\
    \hline
    Historic Set & 2012.9 - 2012.10 & 1544  & 6681 \bigstrut\\
    \hline
    Training Set & 2012.11 - 2013.1 & 2382  & 11494 \bigstrut\\
    \hline
    Test Set & 2013.3 & 672   & 3153 \bigstrut\\
    \hline
    \end{tabular}%
  \label{tab:ds}%
\end{table}%

\section{Evaluation}
\subsection{Experimental Settings}
Dataset. Our datasets are collected through the Twitter stream API with Gardenhose access, containing roughly 10\% of all public statuses on Twitter. We have collected three datasets according to the requirements of the prediction problems. Table~\ref{tab:ds} shows the statistics of these datasets. Historic set is used to find the prototypes of hashtags in training and test sets. Training set is used to training prediction models. Once a prediction model is built, test set is used to evaluate the performance of the model.

Metrics. Since the dataset for prediction task 1 is largely unbalanced, accuracy is not a reasonable metric to evaluate the performance. The classifier can easily get high accuracy by predicting all samples into the class which plays the dominant role in the dataset. Precision, Recall and F-score are reasonable metrics for this prediction task.

\begin{displaymath}
    Precision = \frac{true\_positive}{true\_positive + false\_positive}
\end{displaymath}

\begin{displaymath}
    Recall = \frac{true\_positive}{true\_positive + false\_negative}
\end{displaymath}

\begin{displaymath}
    F_{\beta}-score = (1+\beta^2)\times \frac{Precision\times Recall}{\beta^2Precision + Recall}
\end{displaymath}
Since prediction task 2 and 3 are both regression tasks, the common metric Mean Squared Error($RMSE$) is used to evaluate the prediction performance. The smaller $RMSE$ is, the better the performance is. 

\subsection{Performance of Prediction Task 1}

\begin{table}[!b]
  \centering
  \caption{Performance comparison}
    \begin{tabular}{|c|c|c|c|c|c|c|}
    \hline
    \multirow{2}[4]{*}{Method} & \multicolumn{6}{c|}{$F_1$-score}                 \bigstrut\\
\cline{2-7}          & 5min  & 15min & 30min & 1h    & 3h    & 6h \bigstrut\\
    \hline
    NaiveBayes & 0.299  & 0.342  & 0.312  & 0.197  & 0.132  & 0.081  \bigstrut\\
    \hline
    C4.5  & 0.211  & 0.238  & 0.235  & 0.118  & 0.043  & 0  \bigstrut\\
    \hline
    LR & 0.068  & 0.244  & 0.209  & 0.190  & 0.044  & 0.077  \bigstrut\\
    \hline
    NN & 0.170  & 0.202  & 0.210  & 0.250  & 0.185  & 0.194  \bigstrut\\
    \hline
    SVM   & 0.207  & 0.308  & 0.329  & 0.277  & 0.187  & 0.125  \bigstrut\\
    \hline
    LSM~\cite{nikolov2012trend,chen2013latent}  & 0.184  & 0.145  & 0.161  & 0.097  & 0.014  & 0.022  \bigstrut\\
    \hline
    Our Method & \textbf{0.326}  & \textbf{0.372}  & \textbf{0.368}  & \textbf{0.360}  & \textbf{0.290}  & \textbf{0.250}  \bigstrut\\
    \hline
    \end{tabular} \\
    LR: Logistic Regression \quad NN: Neural Network
  \label{tab:pt1}%
\end{table}%

\subsubsection{Performance Comparison}
Table~\ref{tab:pt1} shows the performance comparison of prediction task 1, bursting hashtag prediction. Predictions were made at six representative time, which can be divided into three stages, early stage (5min, 15min), middle stage (30min, 1h) and late stage (3h, 6h). Predictions after 6 hours are not considered because about 95\% of bursting hashtags get burst within 6 hours since the prediction was triggered. Conclusions can be drawn as follows:
\begin{itemize}
\item Our method significantly outperforms the other related methods in terms of $F_1$-score at any prediction stages. Compared to the best performance of other related methods, the improvement by our method is more significant at later prediction stages. 
\item From 5-minute prediction to 6-hour prediction, the $F_1$-score obtained by any method but LSM increases first and decreases afterwards. It's known to all that the later the prediction is made, the more information can be used for prediction. But why does the performance drop at later prediction stages? The reason is that as time goes by, the dataset become more and more skewed. As shown in Table~\ref{tab:trig}, before 15 minutes, the proportion of positive samples in the dataset is not very low. But after 15 minutes, this proportion becomes lower and lower so that the increase of information used for prediction cannot compensate the proportion decrease of bursting hashtags in the dataset. Therefore the $F_1$-score increases first and decreases afterwards.
\end{itemize}

\begin{figure*}[!b]
\centering
\subfigure[Precision Comparison]{\label{fig:pre}\includegraphics[width=6.0cm, height=4.5cm]{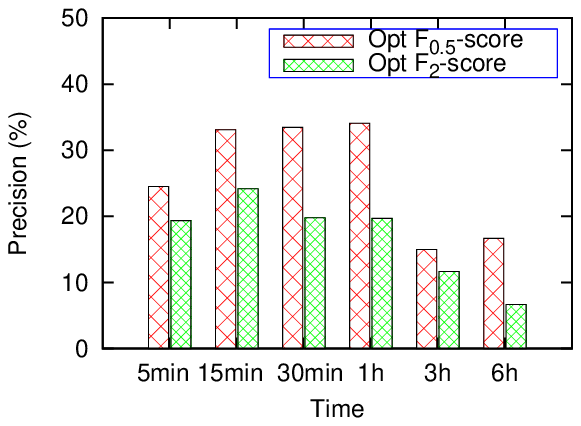}}
\subfigure[Recall Comparison]{\label{fig:rec}\includegraphics[width=6.0cm, height=4.5cm]{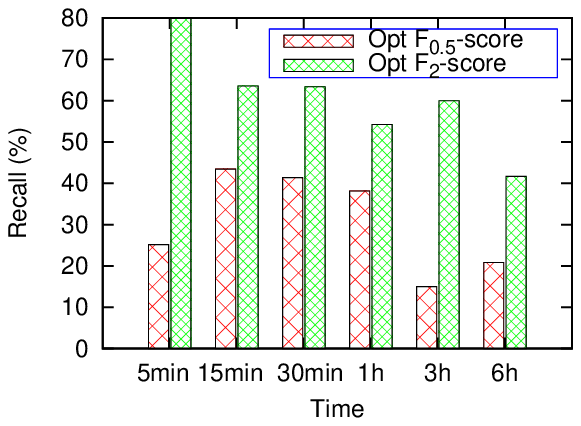}}
\caption{Performance comparison for optimizing prediction models using different F-scores.}
\label{fig:fscmp}
\end{figure*}

In practical applications, there may be different requirements on precision and recall. For a bursting topic recommendation application, high precision seems to be preferred to ensure the quality of recommendation; while for a public opinion monitoring application, high recall tends to be required because it hopes to miss as few bursting topics or events as possible. These requirements can be satisfied by training optimal prediction models using different types of $F$-scores. For example, $F_{0.5}$-score can be used when high precision is preferred, while $F_2$-score can be used when high recall is preferred. Fig.~\ref{fig:fscmp} shows the performance comparison for optimizing prediction models using different $F$-scores. It can be seen that for all the cases, the precision is higher when optimizing $F_{0.5}$-score, while the recall is higher when optimizing $F_2$-score. 

\begin{table*}[!ht]
  \centering
  \caption{Evaluation of Feature Contribution}
    \begin{tabular}{|c|c|c|c|c|c|c|}
    \hline
    \multirow{2}[4]{*}{Feature Removed} & \multicolumn{6}{c|}{$\Delta F_1$-score/$F_1$-score}        \bigstrut\\ 
\cline{2-7}          & 5min  & 15min & 30min & 1h    & 3h    & 6h \bigstrut\\
    \hline
    Meme Features & -2.19\% & 1.95\% & -2.68\% & -4.35\% & \textbf{-18.08\%} & -8.57\% \bigstrut\\
    \hline
    User Features & 1.79\% & 2.76\% & -4.35\% & -6.08\% & \textbf{-41.14\%} & -9.86\% \bigstrut\\
    \hline
    Content Features & 1.74\% & -1.07\% & -4.24\% & -3.18\% & -\textbf{21.34\%} & -12.09\% \bigstrut\\
    \hline
    Network Features & -2.65\% & 2.55\% & -1.88\% & -1.33\% & \textbf{-38.12\%} & -17.76\% \bigstrut\\
    \hline
    Hashtag Features & -1.15\% & 2.79\% & -3.79\% & -13.19\% & -6.37\% & \textbf{-21.31\%} \bigstrut\\
    \hline
    Prototype Features & -3.89\% & 0.71\% & -0.99\% & 0.00\% & -0.72\% & \textbf{-5.88\%} \bigstrut\\
    \hline
    Time Series Features & -5.64\% & -10.30\% & -14.95\% & -32.66\% & -49.40\% & \textbf{-72.41\%} \bigstrut\\
    \hline
    \end{tabular}%
  \label{tab:fc}%
\end{table*}%

\subsubsection{Analysis of Feature Contribution}
We defined and used 7 types of features for predicting bursting hashtags, meme features, user features, content features, network features, hashtag features, time series features and prototype features. In this section, we conduct experiments to examine the contributions of these features to prediction performance. During these experiments, we apply our method a number of times, each time removing only one type of features and recording the changes of prediction performance. The experimental results are demonstrated in Table~\ref{tab:fc}. From Table~\ref{tab:fc}, it can be observed that:
\begin{itemize}
\item Time series features are the most universal and effective features. When they are removed, the drop of $F_1$-score is the most significant for all the cases. Prototype features seems to be the least effective features. For most cases, the drop of $F_1$-score when they are removed is smaller than that when other features are removed. 
\item As time goes by, the contribution of time series features increases progressively. When they are removed, $F_1$-score decreases by 5.64\% for 5-minute prediction; while for 6-hour prediction, the drop of $F_1$-score reaches 72.41\%. The reason is that as time goes by, more and more negative samples show downtrend on time series. When removing any other features, the largest drop of $F_1$-score appears at late prediction stage. This is because the dataset is so skewed at late prediction stage that different types of features are more necessary to be used together to train a synthetic prediction model. 
\end{itemize}

\begin{table*}[!ht]
     \centering
  \caption{Performance for prediction task 2}
    \begin{tabular}{|c|c|c|c|c|c|c|}
    \hline
    \multirow{2}[4]{*}{Method} & \multicolumn{6}{c|}{$RMSE$}                      \bigstrut\\
\cline{2-7}          & 5min  & 15min & 30min & 1h    & 3h    & 6h \bigstrut\\
    \hline
    LR    & 1.397  & 1.390  & 1.664  & 1.593  & 1.504  & 3.238  \bigstrut\\
    \hline
    CART  & 1.413  & 1.413  & 1.675  & 1.761  & 1.618  & 1.295  \bigstrut\\
    \hline
    GPR   & 1.379  & \textbf{1.335}  & 1.570  & \textbf{1.540}  & 0.890  & 1.205  \bigstrut\\
    \hline
    SVR   & \textbf{1.375}  & 1.345  & \textbf{1.566}  & 1.586  & \textbf{0.806}  & \textbf{1.177}  \bigstrut\\
    \hline
    NN    & 1.568  & 1.754  & 2.020  & 2.023  & 1.485  & 1.228  \bigstrut\\
    \hline
    \end{tabular}%
  \label{tab:pt2}%
\end{table*}
\begin{table*}[!hb]
    \centering
  \caption{Performance for prediction task 3}
    \begin{tabular}{|c|c|c|c|c|c|c|}
    \hline
    \multirow{2}[4]{*}{Method} & \multicolumn{6}{c|}{$RMSE$}                      \bigstrut\\
\cline{2-7}          & 5min  & 15min & 30min & 1h    & 3h    & 6h \bigstrut\\
    \hline
    LR    & 1.752  & 1.791  & 1.970  & 1.941  & 1.583  & 1.751  \bigstrut\\
    \hline
    CART  & 2.073  & 1.893  & 1.870  & 1.880  & 1.661  & 1.767  \bigstrut\\
    \hline
    GPR   & \textbf{1.728}  & 1.692  & \textbf{1.711}  & \textbf{1.777}  & 1.376  & 1.691  \bigstrut\\
    \hline
    SVR   & 1.791  & \textbf{1.663}  & 1.721  & 1.789  & \textbf{1.315}  & \textbf{1.672}  \bigstrut\\
    \hline
    NN    & 2.175  & 2.290  & 2.099  & 2.342  & 1.898  & 1.946  \bigstrut\\
    \hline
    \end{tabular}%
  \label{tab:pt3}%
\end{table*}
\subsection{Performance of Prediction Task 2}
For this regression task, we evaluated five different models, Linear Regression, Classification And Regression Tree, Gaussian Process Regression, Support Vector Regression and Neural Network. Table~\ref{tab:pt2} shows the results of performance comparison, measured by the typical metric \emph{RMSE}. It can be seen that SVR gets the best performance for most cases. Only for the 15-minute and 1-hour prediction, is SVR slightly worse than GPR. From the 5-minute prediction to the 6-hour prediction, the performance by SVR drops first and rises afterwards. Obviously, the best performance is achieved at late stage, and the performance of early stage is better than middle stage. This implies that the uncertainty of middle-stage predictions is larger for predicting when a hashtag will burst.


\subsection{Performance of Prediction Task 3}
For this prediction task, we also evaluated the five different models tried in the prediction task 2. Table~\ref{tab:pt3} shows the results of performance comparison. It can be observed that the performance of GPR and SVR is better than other methods. GPR achieves the best performance for 5-minute, 30-minute and 1-hour prediction; while SVR achieves the best performance for the other three cases.


\section{Related Work}
\subsection{Analysis of Hashtag Adoption and Diffusion}
Lin \emph{et al.}~\cite{lin2013bigbirds} studied the dynamics of hashtag adoption by analyzing their growth and persistence. Based on empirical analysis of how the dual role of a hashtag affects hashtag adoption, Yang \emph{et al.}~\cite{yang2012we} predicted whether a user would adopt a hashtag that he never used before. Romero \emph{et al.}~\cite{romero2011differences} studied the diffusion mechanics of hashtags of different types and topics. Chang~\cite{chang2010new} applied diffusion of innovation theory to explain the spread of hashtags on Twitter, i.e., hashtag usage and adoption. Romero~\cite{romero2013interplay} studied the interplay between social relationships and hashtag adoption and found that the social relationships between the initial hashtag adopters can predict future adoption of the hashtag.

\subsection{Popularity Prediction in Microblogging \\Platforms}
Item-level prediction. In microblogging services, interesting tweets are often retweeted by many users. There have been a number of works on popularity prediction of tweets. Suh \emph{et al.}~\cite{suh2010want} performed analytics on factors which may impact retweeting in Twitter. Petrovic \emph{et al.}~\cite{petrovic2011rt} tried to predict whether a new tweet will be retweeted in the future through binary classification. Yang \emph{et al.}~\cite{yang2010understanding} proposed a factor graph model to predict users' retweeting behaviors. Hong \emph{et al.} applied multi-class classification to predict the popularity of tweets~\cite{hong2011predicting}, and used Co-Factorization Machines to address the problem of predicting users' retweeting decisions~\cite{hong2013co}. Kong \emph{et al.}~\cite{kong2012predicting} predicted lifespans of popular tweets based on their static characteristics and dynamic retweeting patterns.

Topic-level prediction. In~\cite{gupta2012predicting}, Gupta \emph{et al.} used regression, classification and hybrid approaches to predict future popularity of current popular events. Given a popular event at the time interval $t$, they predicted the status of this popular event at $t+1$. Analogously, Ma \emph{et al.}~\cite{ma2012will,ma2013predicting} predicted popularity of hashtags in daily granularity. Concretely, they predicted the range of popularity using classification methods. Tsur \emph{et al.}~\cite{tsur2012s} studied the effect of content on the spread of hashtags in weekly granularity. Contrary to these works, this study performs real-time (minute-level) prediction for hashtags. Most germane to this work are two studies from the same group~\cite{nikolov2012trend,chen2013latent}, which focused on predicting trending topics by time series classification. The authors collected some historic trending topics, and selected the slice of time series centered at the trend onset of these topics as reference signals. Then new hashtags were classified by comparing observed time series to these signals. Different from all of these works, our work provides the first definition of bursting hashtags and studies comprehensive real-time prediction problems covering their entire lifecycles.

\section{Conclusion}
In this paper, we take the initiative to propose a real-time prediction framework of bursting hashtags. We define a series of interesting but challenging prediction tasks covering the entire lifecycles of bursting hashtags. Features of different types, as well as solutions are proposed to solve the prediction tasks. Evaluation experiments are conducted on real datasets from Twitter, and the results show that the proposed features and solutions are effective to the prediction tasks.

\end{document}